%% file: EHTran_Main.tex
\documentclass[sigconf]{acmart}

\usepackage{booktabs} 
\usepackage{amsfonts, amsmath,amssymb,amsthm}
\usepackage{algorithm, algorithmic}
\usepackage{array}
\usepackage{cases}
\usepackage{changepage}
\usepackage{epstopdf}
\usepackage[english]{babel}
\usepackage{etoolbox}
\usepackage{graphicx}
\usepackage{multirow}
\usepackage{physics} 	
\usepackage{subfigure}
\usepackage{setspace}
\usepackage{url}  
\usepackage[hyphenbreaks]{breakurl} 
\usepackage{lipsum}

\setcopyright{none} 

\def\ps@headings{%
\def\@oddhead{\mbox{}\scriptsize\rightmark \hfil \thepage}%
\def\@evenhead{\scriptsize\thepage \hfil \leftmark\mbox{}}%
\def\@oddfoot{}%
\def\@evenfoot{}}
\makeatother
\makeatletter
\patchcmd{\@thm}
  {\trivlist}
  {\list{}{\leftmargin=\thm@leftmargin\rightmargin=\thm@rightmargin}}
  {}{}
\patchcmd{\@endtheorem}
  {\endtrivlist}
  {\endlist}
  {}{}
\newlength{\thm@leftmargin}
\newlength{\thm@rightmargin}

\newcommand{\xnewtheorem}[3]{%
  \newenvironment{#3}
    {\thm@leftmargin=#1\relax\thm@rightmargin=#2\relax\begin{#3INNER}}
    {\end{#3INNER}}%
  \newtheorem{#3INNER}%
}

\newtheoremstyle{indentedupright}{3pt}{3pt}{} {}{\bfseries}{.}{.5em}{} 
\newtheoremstyle{indenteditalic}{3pt}{3pt}{\itshape} {}{\bfseries}{.}{.5em}{} 

\theoremstyle{indenteditalic}
\xnewtheorem{0pt}{0pt}{oldpro}{Conventional Property}
\xnewtheorem{0pt}{0pt}{newpro}{New Property}
\xnewtheorem{0pt}{0pt}{obser}{Observation}

\begin{document}
\title{Revisiting Transmission Scheduling in RF Energy Harvesting Wireless Communications}
\author{Yu Luo}
\affiliation{\institution{South Dakota School of Mines and Technology}}
\email{yu.luo@sdsmt.edu}

\author{Lina Pu}
\affiliation{\institution{South Dakota School of Mines and Technology}}
\email{lina.pu@sdsmt.edu}

\author{Yanxiao Zhao}
\affiliation{\institution{South Dakota School of Mines and Technology}}
\email{yanxiao.zhao@sdsmt.edu}

\author{Wei Wang}
\affiliation{\institution{San Diego State University}}
\email{wwang@mail.sdsu.edu }

\author{Qing Yang}
\affiliation{\institution{University of North Texas}}
\email{qing.yang@unt.edu}

\renewcommand{\shortauthors}{Y. Luo et al.}

\input{Abstract.tex}

\settopmatter{printacmref=false} 
\renewcommand\footnotetextcopyrightpermission[1]{} 
\pagestyle{plain} 


\maketitle

\input{Introduction.tex}

\input{RelatedWork.tex}

\input{Primer.tex}

\input{CharFun.tex}

\input{Model.tex}

\input{PerEva.tex}

\input{Conclusion.tex}

\input{Appendix.tex}

\vspace{1cm}

\input{EHTran_Main.bbl}

\end{document}

%% file: Abstract.tex
\begin{abstract}
\label{sec:abstract}
The transmission scheduling is a critical problem in radio frequency (RF) energy harvesting communications. 
Existing transmission strategies in an RF-based energy harvesting system is mainly based on a classic model, in which the data transmission is scheduled in a fixed feasible energy tunnel. In this paper, we re-examine the classic energy harvesting model and show through the theoretical analysis and experimental results that the bounds of feasible energy tunnel are dynamic, which can be affected by the transmission scheduling due to the impact of residual energy on the harvested one. To describe a practical energy harvesting process more accurately, a new model is proposed by adding a feedback loop that reflects the interplay between the energy harvest and the data transmission. Furthermore, to improve network performance, we revisit the design of an optimal transmission scheduling strategy based on the new model. To handle the challenge of the endless feedback loop in the new model, a recursive algorithm is developed. The simulation results reveal that the new transmission scheduling strategy can balance the efficiency of energy reception and energy utilization regardless of the length of energy packets, achieving improved throughput performance for wireless communications.
\end{abstract}

%% file: Introduction.tex
\section{Introduction}
\label{sec:introduction}

Due to the features of self-sustainability,  pollution-free and perpetual operation, energy harvesting becomes a promising technology to drive low-power  devices in future wireless mobile networks ~\cite{huang2015cutting, yang2016heterogeneous, liu2013ambient}. An effective strategy to manage the arrived energy and to schedule the data transmission on an energy harvesting device (EHD) plays a crucial role to achieve a desired network performance in terms of throughput, transmission delay, and communication reliability.

In recent years, many transmission scheduling strategies have been proposed for the radio frequency (RF) based energy harvesting system~\cite{luo2017optimal, xu2014throughput, orhan2014energy, sakulkar2016online}, where both the intermittency and the randomness of energy arrivals are taken into account for the power management. In those designs, the energy grabbed by an EHD has been widely modeled as a random process~\cite{tutuncuoglu2015optimum, gorlatova2013networking, yang2012optimal, tutuncuoglu2012optimum, ozel2011transmission}, which may depend on the activity of RF energy sources (e.g., a TV tower or a cellular base station), but is rarely affected by the transmission scheduling on EHD. In this paper, we challenge this common cognition and confirm through theoretical analysis and  experimental results that the transmission scheduling policy can affect the process of energy harvesting significantly.

Essentially, for a real RF energy harvesting system, there are two different concepts: the \emph{arrived energy} and the \emph{harvested energy}. The former is the efficient energy that reaches an EHD after considering the propagation losses and the power conversion efficiency, while the latter is the energy absorbed and conserved by the EHD's battery. Due to the nonlinear charge characteristic of batteries, the harvested energy is not only determined by the arrived energy but also affected by the residual energy of the EHD~\cite{biason2016effects, gorlatova2013networking}.

\begin{figure}[htb]
\centerline{\includegraphics[width=8.3cm]{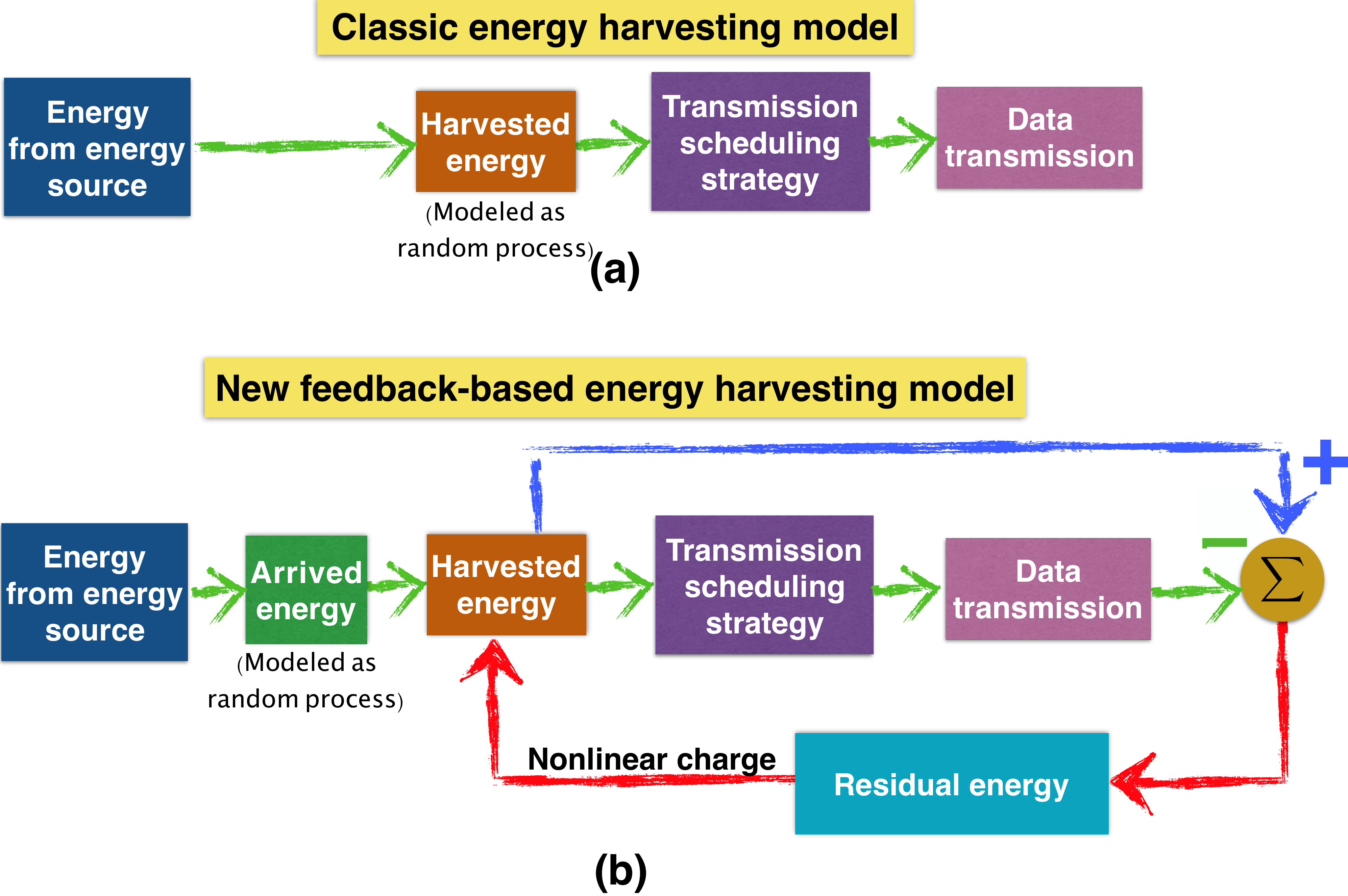}}
\caption{(a) The conventional energy harvesting model, and (b) the new feedback-based model.}\label{fig:ModDif}
\end{figure}

Although the nonlinear charge is a well-known characteristic of batteries, it is not taken into account in the commonly-used energy harvesting model for wireless communications. As illustrated in Fig.\,\ref{fig:ModDif}\,(a), the impact of residual energy of the EHD on the energy harvest process is not considered in the conventional model. With such an assumption, EHDs consuming energy in different manners will harvest an equivalent amount of energy as long as the arrived energy is the same. The existing work in optimal transmission scheduling strategy is to seek a curve within such a fixed energy tunnel so that a pre-defined objective is optimized (e.g., maximizing throughput)\,\cite{ulukus2015energy,ozel2011transmission,tutuncuoglu2012optimum}.

Unfortunately, the above assumption may not be true in real energy harvesting system. Since residual energy is affected by the data transmission, the harvested energy heavily depends on a data transmission strategy. In other words, an EHD cannot estimate what amount of energy it can harvest from an energy packet before scheduling its transmissions, and we call it as the \emph{causality of energy harvest}. The interplay between the transmission scheduling strategy and harvested energy has been completely neglected in the literature. It calls for a new energy harvesting model and re-examination of challenges identified in the existing research.

In this paper, we propose a new energy harvesting model integrating the causality of energy harvest, as illustrated in Fig.\,\ref{fig:ModDif}\,(b). In the proposed model, an energy feedback loop (the red line in Fig.\,\ref{fig:ModDif}\,(b)) from a data transmission to the harvested energy is established that factors the nonlinear charge characteristics of a battery. What distinguishes the classic energy harvesting model is that the feasible energy tunnel is not fixed: its bounds are affected by transmission strategies. Therefore, the formulation of an optimal
transmission strategy based on the new model has to be
revised. With the new feedback-based model, we re-examine the design of an
optimal offline transmission scheduling strategy, which faces grand challenge introduced by the feedback line: on the one hand, the design of an offline transmission scheduling strategy needs to know in advance the amount of energy an EHD can harvest; on the other hand, the transmission scheduling itself affects the energy harvesting process through the residual energy in a battery.

In order to resolve the above challenge, which is referred to as the endless loop problem, we develop a recursive algorithm by leveraging the inherent relationship among the residual energy, transmission power, and harvested energy. In the algorithm, the optimal transmission power in one epoch can be represented by that in prior epochs. Consequently, the design of an optimal transmission scheduling is converted to solving a nonlinear equation. As will be verified in the recursive algorithm, both energies arrived in the past and that will arrive in the future have impacts on the optimal transmission scheduling at the current time.

To summarize, the contributions of our work are threefold. First, through theoretical analysis and experimental results, we verify the limitation of classic energy harvesting model that ignores the interplay between the transmission scheduling strategy and the energy harvesting process in a practical energy harvesting system. Second, a new feedback-based model is proposed, in which the impact of data transmission scheduling on the energy harvest is described via the residual energy.  Third, both an offline optimal and an online suboptimal transmission scheduling strategies are developed using the new feedback-based model. A recursive algorithm is designed to provide an upper bound of throughput that an RF energy harvesting wireless communication system can achieve. Simulation results reveal that compared with existing transmission scheduling strategies, the proposed one can improve the system throughput significantly without violating the causality of energy harvest.

The remainder of this paper is organized as follows. The related work and an overview of the conventional transmission scheduling are introduced in Section~\ref{sec:RelatedWork} and Section~\ref{subsec:primer_model}, respectively. The nonlinear energy harvesting process is discussed in Section~\ref{sec:nonlinear}. The offline optimal transmission scheduling strategy is developed in Section~\ref{sec:Opt}. The simulation results are shown in Section~\ref{sec:perEva} and conclusions are drawn in Section~\ref{sec:conclusion}.

%% file: RelatedWork.tex
\section{Related Work}
\label{sec:RelatedWork}
    
Prior work presented in~\cite{yang2012optimal} proves that, with  the assumption of infinite  battery size, the efficiency of data transmission is maximized when the transmission power remains constant between energy harvests. With such a conclusion, the transmission scheduling at an EHD is simplified into a piecewise-linear optimization problem. Later on, the constraint of battery capacity is taken into account~\cite{ozel2011transmission, tutuncuoglu2012optimum}. Researchers in~\cite{tutuncuoglu2012optimum} construct a feasible energy tunnel, where the upper bound and the lower bound are determined by the constraints of energy causality and battery size, respectively. Geometrically, it is pointed out that to maximize the throughput, the aggregated energy consumption should be the tightest string in the energy feasibility tunnel. The directional water-filling algorithm applied in \cite{ozel2011transmission} is an alternative to optimize data transmission. In such an algorithm, the energy is considered as ``water'', which can neither flow back nor exceed the maximal capacity of a battery, and the algorithm aims at distributing the water equally over time. In \cite{chen2015provisioning}, both the constraints of data and battery capacities are integrated to optimize the transmission scheduling strategy.

To manage the power more efficiently, the energy that an EHD consumes on signal processing is investigated in~\cite{orhan2014energy, xu2014throughput}. As analyzed in \cite{xu2014throughput}, due to the constant overhead of hardware, the energy efficiency of an EHD is non-monotonic with respect to the spectrum efficiency. For the purpose of improving system throughput, a two-phase transmission scheduling policy is provided: the first phase is to maximize the energy efficiency through an on-off power allocation method; the spectrum efficiency is optimized in the second phase with a non-decreasing power allocation strategy. The authors in \cite{orhan2014energy} introduce a directional glue-pouring algorithm to solve a similar problem. Akin to the directional water-filling algorithm, the ``glue'' in \cite{orhan2014energy} is only allowed to flow forward and the equilibrium glue levels are then determined. In addition, due to the constant power consumption over time, a threshold of the transmission power in the directional glue-pouring algorithm needs to be calculated first, and then the process of glue-pouring is performed so that the power level is always higher than the threshold.

Recently, substantial research efforts have been made on realistic scenarios, where  battery imperfections are taken into account~\cite{biason2016effects, devillers2012general, gorlatova2013networking, tutuncuoglu2015optimum}. The authors in \cite{devillers2012general} considers a realistic battery model where the capacity of battery degrades over time and a constant energy leakage incurs. Such imperfections modify the feasible energy tunnel: the distance between the upper and lower bounds monotonously decreases reflecting the time-varying battery capacity. The research conducted in \cite{gorlatova2013networking, tutuncuoglu2015optimum} investigates the charge inefficiencies caused by the nonlinear charge feature of batteries. In the optimal policy design of data transmissions, however, the obtained energy is still modeled as an independent random variables, which neglects the impact of data transmission on  energy harvest. The authors in \cite{biason2016effects} further consider the imperfect knowledge of the instant battery level and propose optimal transmission strategy under limited knowledge (e.g., battery low or hight).

Through an extensive literature survey, we learn that the majority of existing transmission scheduling approaches are based on the classic energy harvesting model, in which the amount of energy replenished is assumed to be random values independent of the way of consuming energy~\cite{devillers2012general, gorlatova2013networking, tutuncuoglu2015optimum, xu2014throughput, orhan2014energy, sakulkar2016online, yang2012optimal, ozel2011transmission, tutuncuoglu2012optimum, chen2015provisioning}. In our paper, we revise the classic model and show through experiments that the harvested energy and the transmission scheduling strategy inherently interact with each other, which must be considered comprehensively for efficient power management.

%% file: Primer.tex
\section{Conventional Scheduling For Data Transmission}
\label{subsec:primer_model}
In this section, we introduce the background knowledge about the conventional transmission scheduling and the corresponding formulation based on the classic harvesting model.

\begin{figure}[htp]
\centering
\subfigure []{
\label{fig:Model_A}
\includegraphics[width=7.8cm]{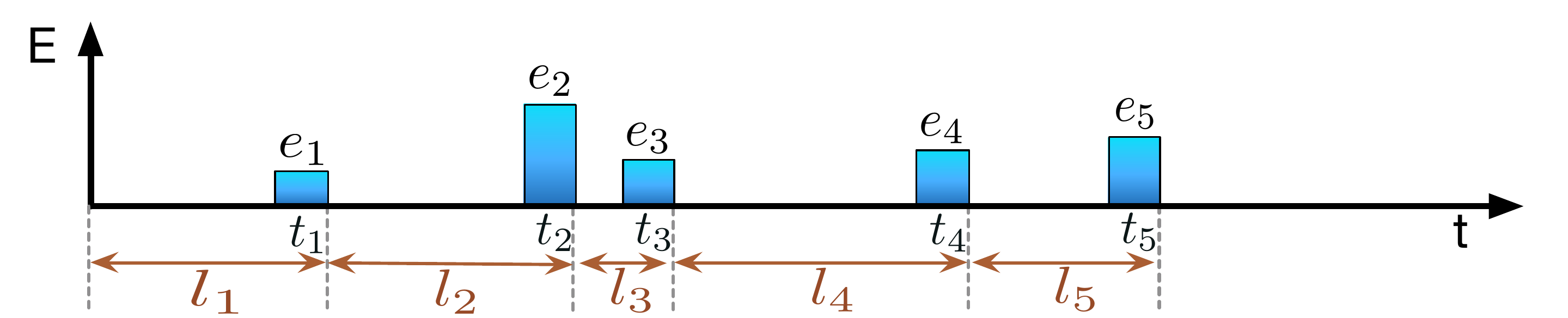}
}%
\vspace{-0.3cm}
\subfigure []{
\label{fig:Model_B}
\includegraphics[width=7.8cm]{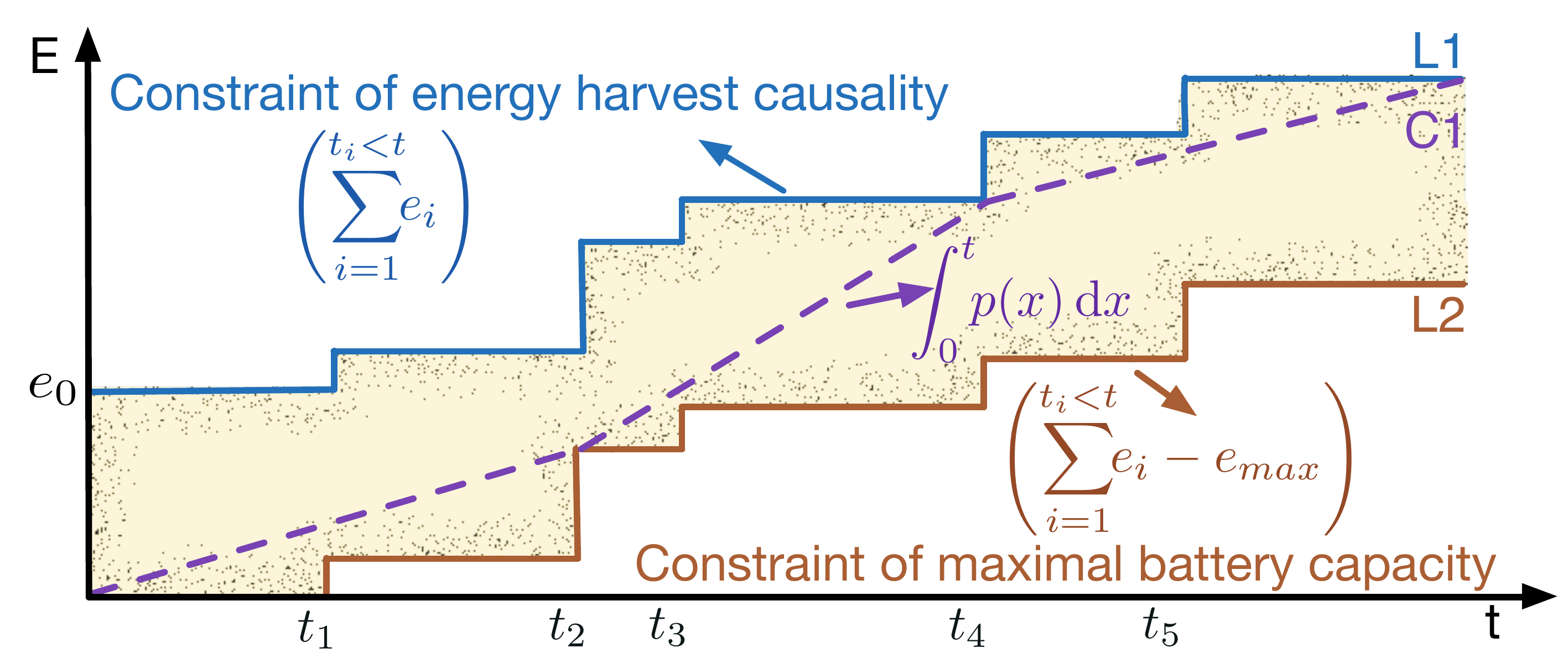}
}%
\vspace{-0.2cm}
\caption{Transmission scheduling based on the classic energy harvesting model. (a) Energy arrival. (b) Feasible energy tunnel.}
\label{fig:Model}
\end{figure}

When an EHD schedules the data transmission, the RF energy is usually considered as discrete energy packets with random sizes, $e_i$, arrived at time, $t_i$, as shown in Fig.\,\ref{fig:Model_A}. The initial energy of the battery is denoted by $e_0$. Assume there are a number of $N$ energy packets transmitted by an energy source in total. The time interval between the successive energy arrivals is called epoch, the length of which is denoted by $l_i$.

Let $p(t)$ be the instantaneous transmission power of an EHD at time $t$. According to the energy causality constraint, the EHD cannot use the energy that has not arrived yet, i.e.,
\begin{equation}
\label{eq:3_01}
	\int_0^{t }p(x)\,\mathrm{d}x\leq\sum_{i=0}^{{t_i<t}} e_i.
\end{equation}
In addition, let $e_m$ be the capacity of battery, which is the maximum energy can be stored in the EHD. Due to the constraint of battery size, it yields:
\begin{equation}
\label{eq:3_02}
	\sum_{i=0}^{{t_i<t}} e_i -\!\int_0^{t }p(x)\,\mathrm{d}x\leq e_m;
\end{equation}
otherwise the energy overflow occurs.

As shown in Fig.\,\ref{fig:Model_B}, the constraints expressed in (\ref{eq:3_01}) and (\ref{eq:3_02}) define an upper bound and a lower bound for energy conservation, respectively. In other words, the profile of the aggregated energy consumption, namely, $\int_0^{t }p(x)\,\mathrm{d}x$, must stay within the energy tunnel constrained by (\ref{eq:3_01}) and (\ref{eq:3_02}). 

Let $r(t)$ be the transmission rate of an EHD at time $t$, which is related to the transmission power, $p(t)$, through a power-rate function, $r(t)=\mathcal{G}(p(t))$.  In general, $\mathcal{G}(\cdot)$ is increasing, non-negative, and strictly concave. In real applications, an energy harvesting system usually works in an additive white Gaussian noise (AWGN) channel. Consequently, the power-rate function is $r(t)\!=\!\log_2[1+h(t)p(t)]$, where $h(t)$ is the instantaneous channel response between the EHD and its intended receiver. The goal of an optimal transmission scheduling strategy is to  maximize the throughput subject to the constraints of energy causality in (\ref{eq:3_01}) and maximum battery capacity in (\ref{eq:3_02}).

The majority of existing transmission scheduling policies are based on the classic energy harvesting model. They assume that the energy captured in each epoch, i.e., $e_i$, is a random value, which is only determined by the activity of energy source, the propagation loss of radio waves, and the power conversion efficiency of EHDs. Under this assumption, the transmission scheduling strategy does not affect the amount of energy harvested by the EHD in each epoch.

The optimal transmission power strategy based on the classic energy harvesting model has the following two critical properties:
\begin{oldpro}
\label{oldpro:01}
	\emph\emph{{The transmission power will not change until the battery is either full or completely depleted.}}
\end{oldpro}
\vspace{0.01cm}
\begin{oldpro}
\label{oldpro:02}
	\emph\emph{{The transmission power decreases or increases only at energy arrival instants when the battery is full or completely depleted, respectively.}}
\end{oldpro}
\noindent

In this paper, we will prove that the classic model and its conventional properties do not match the energy harvesting process in a real system very well, which motivates us to develop a more realistic energy harvesting model and to re-examine the issue of optimal transmission scheduling in the new model.

%% file: CharFun.tex
\section{Nonlinear Energy Harvesting Process}
\label{sec:nonlinear}

As depicted in Fig.~\!\ref{fig:ModDif}, the new energy harvesting model is featured by its feedback line that displays the interplay between the transmission scheduling strategy and the energy harvesting process through a nonlinear charge function of a battery. In this section, the charge characteristic of an EHD is briefly introduced with theoretical analysis and  experimental measurements.

\subsection{Charge Characteristic of EHDs}
\label{subsec:charge}
We first theoretically analyze the charge characteristic of an EHD, in which a supercapacitor is commonly used as a battery ~\cite{pinuela2013ambient}. The amount of energy arrived at EHDs depends on the power density, i.e., the strength of RF signals, and the duration of energy packets. The power density affects the highest voltage that the supercapacitor can be charged to~\cite{jabbar2010rf}. According to the measurements reported in \cite{bouchouicha2010ambient}, in the urban environment, the densities of ambient RF signal on different frequency bands (680\,MHz-3.5\,GHz) are almost constants over time. Therefore, the variation of arrived energy in our model is mainly factored by the length of energy packet $i$, which is denoted by $T^e_i$.

Assume the initial energy of the capacitor is $e_0\!=\!0$. Let $V^r_i$ and $E^r_i$ be the voltage and the residual energy of the capacitor at the instant of energy packet $i$ arrived, respectively. The voltage increase of the capacitor is denoted by $\Delta v_i$ with $E^h_i$ joules of energy harvested from energy packet $i$. The resistance of the RC charging circuit and the capacitance of the supercapacitor are represented by $R$ and $C$, respectively. The time constant of the RC charging circuit is denoted by $\tau$, where $\tau\!=\!R\,C$. Let $e_m=\frac{1}{2}Cv_m^2$ be the highest capacity that the supercapacitor can reach, where $v_m$ is the highest voltage the capacitor can be charged in the current RF environment. According to the charge characteristic of the capacitor, we have that:
\begin{equation}\label{eq:4_01}
    \left\{
    \begin{array}{lll}
     \vspace{0.05cm}
        V^r_i = v_m\left(1-e^{-\frac{t^r_i}{\tau}}\right), \\
        \vspace{0.05cm}
        V^r_i+\Delta v_i = v_m\left(1-e^{-\frac{t^r_i+T^e_i}{\tau}}\right),\\ 
        \vspace{0.1cm}
        E^r_i= \displaystyle\frac{1}{2}C\left(V^r_i\right)^2,\\
        \vspace{0.05cm}
        E^r_i+E^h_i = \displaystyle\frac{1}{2}C(V^r_i+\Delta v_i)^2.
    \end{array}
    \right.
\end{equation}
In the above equation set,  $t^r_i$ is the aggregated time spent on charging the voltage of a capacitor from 0 to $V^r_i$. By solving (\ref{eq:4_01}), the amount of harvested energy is calculated as below:
\begin{equation}\label{eq:4_06}
E^h_i= A^1_i(A^2_i)^2\!+\!A^1_iA^3_i(E^r_i)^\frac{1}{2}\!+\!A^1_iA^4_iE^r_i,
\end{equation}
where 
\begin{equation}\label{eq:4_07}
\begin{aligned}
    A^1_i &= \displaystyle\frac{1}{2}e^{-\frac{2T^e_i}{\tau}}, \;\; A^2_i = (2e_m)^\frac{1}{2}\left(e^{\frac{T^e_i}{\tau}}-1\right), \\
  A^3_i &= 2^\frac{3}{2}A^2_i\;, \;\;\;\;\, A^4_i = 2\left(1-e^{\frac{2T^e_i}{\tau}}\right).
  \end{aligned}
\end{equation}

Equation (\ref{eq:4_06}) describes the dependence of  the harvested energy on both the arrived energy (i.e., in terms of $T^e_i$ and $v_m$) and the residual energy (i.e., $E^r_i$). Furthermore, by inspecting the first order and the second order derivatives of $E^h_i$ with respect to $E^r_i$, it can be verified that $E^h_i$ is a concave function of $E^r_i$, since $A^1_i$ and $A^3_i$ are positive  and $A^4_i$ is negative. After calculations, it can be obtained that $E^h_i$ reaches the maximum (i.e., $E^h_{i(max)}$) when
\begin{equation}
\label{eq:4_08}
    E^r_{i(max)} = \displaystyle\frac{e_m}{\left(1+e^{T^e_i/\tau}\right)^2},\quad i=1,\ldots,N+1.
\end{equation}
According to (\ref{eq:4_08}), an EHD can maximize the received energy by retaining a specific amount of energy depending on the length of energy packet. However, it is worth noting that more harvested energy does not guarantee a higher throughput. As will be revealed in Section~\ref{subsec:perEva_onOff}, an optimal transmission scheduling strategy needs to balance the efficiency of energy harvest and energy utilization.

\subsection{Experimental Measurements}
\label{subsec:Exp}

\begin{figure}[htp]
\centerline{\includegraphics[width=6.0cm]{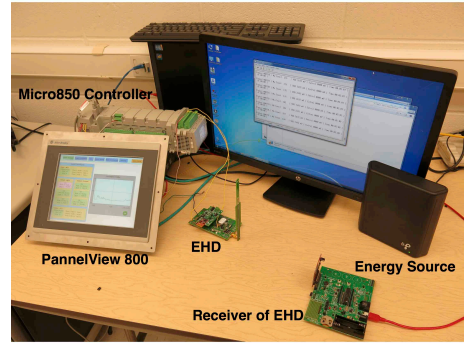}}
\caption{Experimental setup, where EHD grabs RF energy radiated from the energy source to power the sensor and its wireless communication module. The computer connected with a receiver  records the sensing data sent from the EHD on 2.4 GHz frequency band.}\label{fig:exp_setup}
\end{figure}

To evaluate the nonlinear relationship between the harvested energy and the residual energy, an indoor experiment  was conducted using Powercast P2110 energy harvesting kit~\cite{powercast2016powercast},   Micro850 controller and PannelView800 display, as shown in Fig.~\ref{fig:exp_setup}.

In the test, an EHD with $1$\,dBi omnidirectional antenna was scheduled to receive the RF energy radiated from a dedicated energy source, TX91501 transmitter, which was placed about $6$ ft away from the EHD. The effective isotropic radiated power and the frequency of the energy transmitter were $3$\,W and $915$\,MHz, respectively. In order to reduce the energy dissipation, a supercapacitor with low equivalent series resistance (ESR) was selected as the battery. Since TX91501 is non-programmable, the ON/OFF time of TX91501 was manipulated via Micro850 to control the charging time, $T^e$. The controller calculated the energy harvested by the EHD and displayed the results on PannelView800.

\begin{figure}[htb]
\centerline{\includegraphics[width=7.0cm]{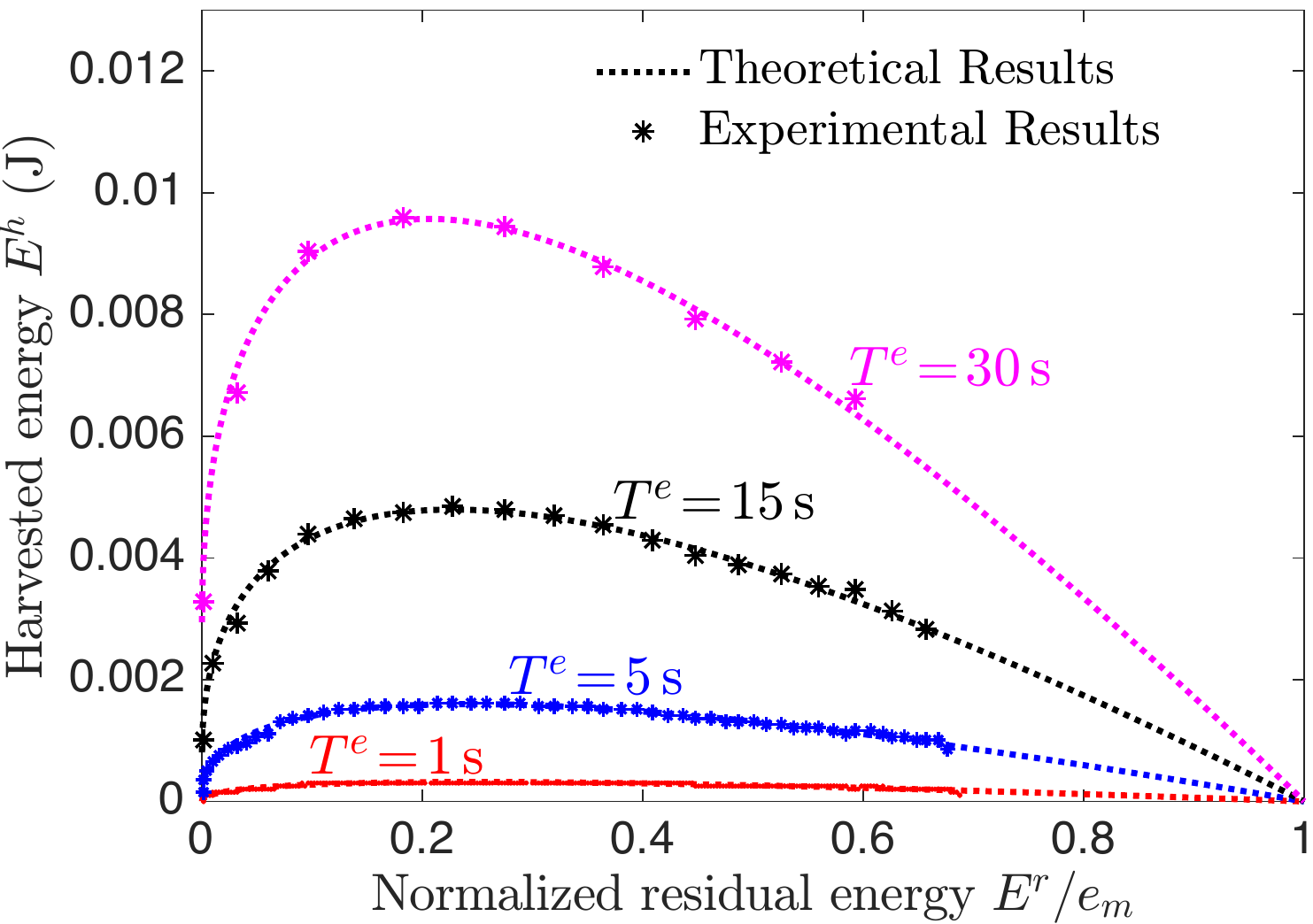}}
\caption{Harvested energy changes nonlinearly with respect to $E^r/e_m$, where $C=0.68$\,F, $R=750$\,$\Omega$, $v_m=2.5$\,V.}\label{fig:charEng}
\end{figure}

In Fig.\,\ref{fig:charEng}, we show how the harvested energy changes nonlinearly with the normalized residual energy for different lengths of energy packet (i.e., $T^e$). How the figure, it can be observed that $E^h$ is not an independent value that assumed in the classic energy harvesting model, but significantly relies  on $E^r$. Moreover, as demonstrated in the figure, $E^h$ is a non-monotonic concave function of $E^r$. In other words, there exists a global optimal $E^r$ for the highest energy harvest efficiency. Using $T^e\!=\!15$\,s as an example, an EHD can capture at most $4.8$\,mJ of energy when $E^r\!\approx\!0.23\times e_m$ in the experiment, which matches the result calculated from (\ref{eq:4_08}) very well.

%% file: Model.tex
\section{New Transmission Scheduling}
\label{sec:Opt}

In this section, we investigate an offline optimal transmission scheduling to maximize the throughput of RF energy harvesting systems following the new feedback-based energy harvesting model. Compared to solutions from the classic model, a primary innovation and challenge are to establish a correct connection between the harvested energy and the transmission strategy. The problem of optimal transmission scheduling is reformulated with two new constraints and eventually solved with a recursive algorithm.

\subsection{Transmission Scheduling in New Model}
\label{subsec:Opt}
In the energy harvesting wireless communications, the relationship between residual and harvested energy can be expressed as:
\begin{equation}
\label{eq:5_01}
    E^r_i = \sum_{j=0}^{i-1} E^h_j-\int_0^{t _i}p(t)\,\mathrm{d}t, \quad i=1,\ldots,N\!+\!1,
\end{equation}
where $E^h_0\!=\!e_0$ is the initial energy of the supercapacitor. The two terms in the right-hand side of (\ref{eq:5_01}) represent the cumulation of energy harvesting and energy consumption, respectively. Combining with the relationship between $E^h_i$ and $E^r_i$ indicated in (\ref{eq:4_06}), $E^h_i$ in each epoch can be recursively represented by $p(t)$, i.e.,
\begin{equation}\label{eq:5_02}
        E^h_i = \mathcal{Q}_i\!\left(\,p(t)\right), \quad 0\leq t<t_i,\;\;i=1, \ldots,N.
\end{equation}
Here, we call $\mathcal{Q}(\cdot)$ as the power-harvest function, which indicates the impact of transmission scheduling, $p(t)$, on harvested energy, $E^h_i$. This relationship indicates that an EHD cannot estimate the amount of energy it can harvest from an energy packet before scheduling its data transmissions, and we call it as the \emph{causality of energy harvest}. 

The power-harvest function reveals the inherent relationship between $p(t)$ and $E^h_i$, which allows for a more realistic and accurate description of energy harvesting process compared with the classic model. By integrating the power-harvest function, the energy causality constraint in (\ref{eq:3_01}) and the battery capacity limit in (\ref{eq:3_02}) are rewritten as (\ref{eq:5_03}) and (\ref{eq:5_04}), respectively.
\begin{numcases}{}
        E^h_0\!+\!\displaystyle\sum_{j=1}^{i-1}\!\mathcal{Q}_j\!-\!\!\int_0^{t _i}\!\!p(t)\,\mathrm{d}t\geq0, \; i=1,\ldots,N\!+\!1.\label{eq:5_03}\\
\vspace{-0.2cm}
        E^h_0\!+\!\displaystyle\sum_{j=1}^{i}\mathcal{Q}_j\!-\!\!\int_0^{t _i}\!p(t)\,\mathrm{d}t\leq e_m, \;\; i=1,\ldots,N.\label{eq:5_04}
\end{numcases}

Based on the above two new constraints, designing an optimal transmission scheduling strategy to maximize the throughput of an EHD is formulated as follows:
\begin{equation}\label{eq:5_05}
\begin{array}{lll}
\hspace{-0.36cm}
    \textbf{\emph{P1}}\;\;
    \vspace{0.1cm}
    \underset{p(t)}{\mathrm{arg\,max}}\displaystyle\int_0^{t_{N\!+\!1}} \mathcal{G}\left(\,p(t)\right)\,\mathrm{d}t, \\
    \begin{array}{lll}
    \vspace{0.0cm}
    \hspace{-0.6cm}
    \textbf{s.t.}\\

    \vspace{0.0cm}
    \hspace{-0.5cm}
    \textbf{\emph{C1}}
            \displaystyle\int_0^{t _i}\!\!p(t)\,\mathrm{d}t\!-\!\sum_{j=1}^{i-1}\!\mathcal{Q}_j\!-\!E^h_0\leq0, \!\!\!&i&\!\!\!\!\!=1,\ldots,N\!+\!1,\\

    \hspace{-0.5cm}
    \textbf{\emph{C2}}\;\; E^h_0\!+\!\displaystyle\sum_{j=1}^{i}\mathcal{Q}_j\!-\!\!\int_0^{t _i}\!p(t)\,\mathrm{d}t\leq e_m, \!\!&i&\!\!\!\!=1,\ldots,N.
    \end{array}
    \end{array}
\end{equation}

The new optimization problem \emph{P1} differs from the conventional one with two distinct constraints, \emph{C1} and \emph{C2}. In those two constraints, the interplay between the harvested energy (i.e., $E^h$) and the data transmission policy (i.e., $p(t)$) is fully incorporated in the power-harvest function, $\mathcal{Q}(\cdot)$. This makes the new transmission scheduling more realistic than existing strategies, which usually do not obey the constraint of energy harvest causality.

\subsection{Optimal Offline Transmission Scheduling}
\label{subsec:offopt}
In the new energy harvesting model, the feasible energy tunnel, in which data transmission is scheduled, is not fixed anymore but varies with different transmission strategies. In such a tunnel, the transmission policy and the harvested energy interact with each other, which causes an endless loop problem. Next, we introduce how to utilize inherent features in \emph{P1} to solve the problem.

Referring to the proof of  Lemma 2 in \cite{yang2012optimal}, it can be proved that the optimal transmission power, which is denoted by $p^*(t)$, is a piecewise linear function. 
Therefore, let $p_i$ be the EHD's transmission power at epoch $i$, then the residual energy in (\ref{eq:5_01}) can be rewritten as:
\begin{equation}
\label{eq:6_01}
    E^r_i = \sum_{j=0}^{i-1} E^h_j- \sum_{j=1}^{i}p_j l_j, \quad i=1,\ldots,N\!+\!1.
\end{equation}
Using (\ref{eq:6_01}), $\mathcal{Q}_i\!\left(\,p(t)\right)$ in (\ref{eq:5_02}) can then be represented as $\mathcal{\tilde{Q}}_i\!\left(p_1,\ldots,p_i\right)$.

According to the piecewise linear feature of the optimal transmission scheduling strategy and (\ref{eq:6_01}), \emph{P1} is converted into the following optimization problem in an AWGN channel:
\begin{equation}\label{eq:6_02}
\begin{array}{lll}
    \hspace{-0.36cm}
    \textbf{\emph{P2}}\;\;
    \vspace{-0.1cm}
    \underset{p_i}{\mathrm{arg\,max}}\displaystyle\sum^{N+1}_{i=1}\frac{l_i}{2}\log_2(1+p_i), \\
    \begin{array}{lll}
    \vspace{-0.15cm}
    \hspace{-0.5cm}
    \textbf{s.t.}\\

    \vspace{0.0cm}
    \hspace{-0.5cm}
    \textbf{\emph{C1}}\;\;\,
            \displaystyle\sum_{j=1}^{i}p_j l_j\!-\!\sum_{j=1}^{i-1}\!\mathcal{\tilde{Q}}_j\!-\!E^h_0\leq0, &i&\!\!\!\!= 1,\ldots,N\!+\!1,\\

    \hspace{-0.5cm}
    \textbf{\emph{C2}}\;\;\, E^h_0\!+\!\displaystyle\sum_{j=1}^{i}\mathcal{\tilde{Q}}_j\!-\!\!\sum_{j=1}^{i}p_j l_j\leq e_m, \,\; &i&\!\!\!\!=1,\ldots,N.
    \end{array}
    \end{array}
\end{equation}

From (\ref{eq:6_02}), it can be observed that \emph{P2} is a nonlinear maximization problem with inequality constraints, and  Karush-Kuhn-Tucker (KKT)  are the first-order necessary conditions in \emph{P2} for the optimal solution. Obviously, the objective function is concave as its Hessian matrix is negative semidefinite for all $p_i$. However, $\mathcal{\tilde{Q}}_j$ is nonlinear and complex with respect to $p_i$ for $i=1\ldots,N$. Consequently, it is hard to identify whether the constraints  $C1$ and $C2$ are convex or not. Therefore, KKT may not be sufficient conditions and multiple  solutions may exist in KKT. Hence, we substitute each solution into \emph{P2} and select the one that maximizes the objective function subject to the constraints as the optimal strategy, which is denoted by $p^*_i$. 

Let $\{\lambda_i\}$ and $\{\mu_i\}$ be the KKT multipliers respectively associated with the constraints $C1$ and $C2$ of \emph{P2}. From the classic optimization theory, the KKT conditions of \emph{P2} are enumerated as follows:

\vspace{0.15cm}
\hspace{-0.35cm}
\textbf{Stationarity:}
\begin{equation}\label{eq:6_03}
\hspace{-0.05cm}
\begin{array}{lll}
    \vspace{0.05cm}
   \nabla_{\!p^*_m}\mathcal{L}\!\!\!\!\!\!&=\!\nabla_{\!p^*_m}\!\!\left(\displaystyle\sum^{N+1}_{i=1}\frac{l_i}{2}\log_2(1\!\!+\!p^*_i)\!\!\right)\\

    \vspace{0.05cm}
    &-\!\!\displaystyle\sum_{i=1}^{N+1}\lambda_i\,\nabla_{\!p^*_m}\!\!\left(\,\sum_{j=1}^{i}p^*_j l_j\!-\!\!\sum_{j=1}^{i-1}\!\mathcal{\tilde{Q}}_j\!-\!E^h_0\!\right)\\

    \vspace{0.05cm}
    &-\!\displaystyle\sum_{i=1}^{N}\!\mu_i\,\nabla_{\!p^*_m}\!\!\left(\!E^h_0\!+\!\!\displaystyle\sum_{j=1}^{i}\mathcal{\tilde{Q}}_j\!-\!\!\sum_{j=1}^{i}p^*_j l_j\!-e_m\!\right) =0,\\
    \end{array}
\end{equation}
where $\mathcal{L}$ is the Lagrangian depending on $p_i$, $\lambda_i$ and $\mu_i$; $\nabla\!_x(\cdot)$ represents the partial derivative with respect to $x$, and $m\!=\!1, \ldots ,N\!+\!1$.

\vspace{0.15cm}
\hspace{-0.35cm}
\textbf{Primal feasibility:}
\begin{equation}\label{eq:6_04}
    \left\{
    \begin{array}{lll}
\displaystyle\sum_{j=1}^{i}p^*_j l_j\!-\!\sum_{j=1}^{i-1}\!\mathcal{\tilde{Q}}_j\!-\!E^h_0\leq0,&i=1,\ldots,N\!+\!1,\\
E^h_0\!+\!\displaystyle\sum_{j=1}^{i}\mathcal{\tilde{Q}}_j\!-\!\!\sum_{j=1}^{i}p^*_j l_j\leq e_m,&i=1,\ldots,N.
    \end{array}
    \right.
\end{equation}

\hspace{-0.35cm}
\textbf{Dual feasibility:}
\begin{equation}\label{eq:6_07}
    \left\{
    \begin{array}{lll}
\lambda_i\geq0, &i=1,\ldots,N\!+\!1,\\
\mu_i\geq0, &i=1,\ldots,N. \\
    \end{array}
    \right.
\end{equation}

\hspace{-0.35cm}
\textbf{Complementary slackness:}
\begin{numcases}{\!\!\!\!\!\!\!\!\!\!}
\!\!\!\lambda_i\left(\,\sum_{j=1}^{i}p^*_j l_j\!-\!\sum_{j=1}^{i-1}\!\mathcal{\tilde{Q}}_j\!-\!E^h_0\!\right)\!=\!0, \!\!\!\!\!\!\!\!\!\!\!&$i=1,\ldots,N\!+\!1$, \label{eq:6_10}\\
\!\!\!\mu_i\left(E^h_0\!+\!\displaystyle\sum_{j=1}^{i}\mathcal{\tilde{Q}}_j\!-\!\sum_{j=1}^{i}p^*_j l_j\!-\!e_m\!\right)\!=\!0, \!\!\!\!\!\!\!\!\!&$i=1,\ldots,N$. \label{eq:6_11}
\end{numcases}

Next, how to calculate the four KKT conditions to find all potential $p^*_i$ is discussed. We first introduce two new properties of the optimal transmission scheduling strategy, which can be used to simplify the complementary slackness of the KKT conditions.
\begin{newpro}
\label{newpro:01}
    \emph\emph{{The new optimal transmission scheduling strategy will not completely deplete the battery until all data are transmitted, i.e., $\forall i\!\in\!\mathbb{Z}^+, i\leq N\!+\!1\!\!:\,\! E^r_i\!>\!0$, unless  $i\!=\!N\!+\!1$.}}
\end{newpro}

\begin{newpro}
\label{newpro:02}
    \emph\emph{{The new optimal transmission scheduling strategy will not fully charge an EHD, i.e., $\forall i\!\in\!\mathbb{Z}^+, i\leq N\!+\!1\!\!:\,\! E^r_i\!+\!E^h_i\!<\!e_m$.}}
\end{newpro}

\begin{adjustwidth}{0cm}{0cm}
The proof of New Property~\ref{newpro:01} can be found in Appendix~\ref{app:appA}. It emphasizes that in the new optimal transmission strategy, an EHD will retain a positive amount of residual energy until the end of data transmission, i.e., $E^r_{N+1}\!=\!0$. This feature will be used in Appendix~\ref{app:appB} to solve the KKT conditions. New Property~\ref{newpro:02} is a result of the exponential component of the charging function presented in (\ref{eq:4_01}). The charging current approaches zero when a battery's voltage is close to the maximal value and it will take infinite time to fully charge the EHD. Comparing with the Conventional Property~\ref{oldpro:01} and \ref{oldpro:02} introduced in Section\,\ref{subsec:primer_model}, which  require an EHD to fully charge or completely deplete the battery before the change of transmission power, the new properties of an optimal strategy fit the real feature of energy harvesting module much better. 
\end{adjustwidth}

Now, we solve the KKT conditions starting from the stationarity equations. Through the derivation performed in Appendix~\ref{app:appB}, the optimal transmission power of an EHD in the current epoch can be represented by that in previous epochs through the following iteration expression:
\begin{equation}
\label{eq:app_13}
 p^*_{m+1}=\left(1+p^*_m\right)(X_m+1)-1,
\end{equation}
where $X_m=\frac{1}{2}A^1_mA^3_m\left(E^r_m\right)^{-\frac{1}{2}}\!+\!A^1_mA^4_m$, and $A^i_m$ ($i=1,3,4$) has been listed in (\ref{eq:4_07}). Given the iteration, a recursive algorithm, which is referred to as Algorithm~\ref{alg:6_01}, can be developed to calculate the optimal transmission scheduling.

\begin{algorithm}[htb]
\caption{Calculation of the optimal transmission power}
\label{alg:6_01}
\begin{algorithmic}[1]
    \STATE Let $E^r_1$ be the single variable.
    \STATE Represent $p^*_1$ by $E^r_1$ based on $p^*_1=\frac{E^h_0-E^r_1}{l_1}$. \label{alg:01}
    \vspace{0.12cm}
    \FOR{$i=1$ to $N$} \label{alg:05}
    \vspace{0.07cm}
        \STATE Represent $E^h_{i}$ by $E^r_1$ based on (\ref{eq:4_06}).
    \vspace{0.12cm}
    \STATE Represent $p^*_{i+1}$ by $E^r_1$ based on (\ref{eq:app_13}).
    \vspace{0.12cm}
    \STATE Represent $E^r_{i+1}$ by $E^r_1$ based on (\ref{eq:6_01}).
    \vspace{0.07cm}
    \ENDFOR \label{alg:03}
    \vspace{0.12cm}
    \STATE To fully utilize all conserved energy by the end of data transmissions, we set $E^r_{N+1}\!=\!0$, and then calculate $E^r_1$, which may have multiple solutions.\label{alg:04}
    \vspace{0.12cm}
    \STATE Calculate all potential $p^*_i$, $i=1,\ldots,N+1$, by substituting all solutions of $E^r_1$ into Step \ref{alg:01} to Step \ref{alg:04}.
    \vspace{0.12cm}
    \STATE Substituting all potential  $p^*_i$ into \emph{P2}, the one that maximizes the objective function subject to the constraints is selected as the optimal strategy.
\end{algorithmic}
\end{algorithm}

By executing commands between Step 2 and Step 7 of Algorithm~\ref{alg:6_01}, $E^r_2$ to $E^r_{N+1}$ are replaced by $E^r_1$ iteratively. As a consequence, according to the New Property~\ref{newpro:01} that $E^r_{N+1}\!=\!0$, a nonlinear equation, $\mathcal{J}(E^1_r)$, with the single variable, $E^1_r$, is available at Step 8. Through using a numerical approach like Newton-Raphson, $E^1_r$ can be calculated from $\mathcal{J}(E^1_r)$. Eventually, all potential optimal transmission power, $p^*_i$, in each epoch is determined iteratively according to the relationship between the transmission power, residual energy, and harvested energy described in Step 2 to Step 7 of the algorithm.

It is worth noting that after the recursive process of Algorithm~\ref{alg:6_01}, $\mathcal{J}(E^1_r)$ contains the coefficients from $A^1_i$ to $A^4_{i}$, where $i\!=\!1, \ldots, N$. According to (\ref{eq:4_07}), $A^x_i$ is correlated with the length of energy packet $i$. This indicates that the optimal transmission power of an EHD in current epoch is affected not only by the energy received in previous epochs but also by the future energy arrivals, which poses a grand challenge on the development of an optimal online transmission scheduling strategy. In Section~\!\ref{subsec:perEva_online}, we will have a short discussion on the design of a suboptimal online strategy by using the method of energy prediction. 

Although the new transmission scheduling strategy developed in this section can only work in an offline manner, it is still valuable in real applications because:
\vspace{0.1cm}
\begin{adjustwidth}{0.1cm}{0cm}
\begin{description}
\setlength{\labelsep}{0.3em}
\itemsep 0.07cm
  \item[a)] It provides an upper bound of the throughput that an RF energy harvesting communication system can reach.
  \item[b)] The new offline strategy can be applied to an online scenario with energy prediction approaches. As reported in \cite{bouchouicha2010ambient}, the energy densities of ambient RF signal on different frequency bands (680\,MHz-3.5\,GHz) are almost constants over time in the urban environment, providing a possibility to estimate the future energy arrivals accurately. 
\end{description}
\end{adjustwidth}

%% file: PerEva.tex
\section{Simulation Results}
\label{sec:perEva}

In this section, we evaluate the proposed offline optimal transmission scheduling and the conventional strategies through simulations.
As discussed in \cite{ulukus2015energy}, conventional strategies assume the data transmission and the energy harvesting process are independent, which violates the \emph{causality of energy harvest} introduced in Section~\ref{subsec:Opt}. In order to make a feasible comparison in simulations, we generate the energy tunnel first and then calculate the corresponding lengths of arrived energy  based on the specific $E^h$ and $E^r$, which is in reverse order of a real energy harvesting process. 

\subsection{Channel Model and Simulation Settings}
\label{subsec:perEva_set}

In the simulation, we use a common AWGN channel model. The noise power denoted by $N_l$ is calculated through $N_l\!=\!N_0+10\log_{10} (B)$, where $B$ is the communication bandwidth and $N_0\!=\!-174$\,dBm is the noise density. The distance between EHD and its intended receiver, $d$, is $10$\,ft, and the propagation loss of RF signals from EHD to the receiver is calculated through the free space path loss (FSPL) model, where
\begin{equation}\label{eq:9_01}
    \text{FSPL\,(dB)}=20\log_{10} (d)+20\log_{10} (f)-147.55.
\end{equation}
According to the capacity of AWGN channel and propagation loss model, the power-rate function on a unit bandwidth is
\begin{equation}\label{eq:eh20}
    r(t)\!=\!\log_2\left[1+p(t)-\text{FSPL}-N_l\right],
\end{equation}
where $r(t)$ is the data transmission rate in bps and $p(t)$ is the transmission power in dBm.
The central frequency and bandwidth for data transmission are $2.4$\,GHz and $10$\,MHz, respectively. Moreover, the resistance of the EHD's charging circuit, $R$, is $4$\,k$\Omega$ and the capacitance of the capacitor, $C$, is $50$\,mF.

For comparison purpose, the performance of the maximal energy harvesting strategy obtained through (\ref{eq:4_08}) and the conventional policy proposed in \cite{tutuncuoglu2012optimum} are assessed. The maximal energy harvesting strategy aims at maximizing the energy harvest each time but ignores the efficiency of energy utilization. To prevent the conventional strategy from violating the causality of energy harvest, we implement the energy tunnel as follows:

\vspace{0.1cm}
\begin{adjustwidth}{0cm}{0cm}
\begin{itemize}

  \item[\textendash] \emph{Step 1:} Generate a random energy tunnel for the conventional strategy, where $(E^h_i)^\prime$ represents the energy harvested from energy packet $i$. Apply the conventional strategy to schedule the data transmission\footnote{According to the New Property \ref{newpro:02}, an EHD cannot be charged to $e_m$ within a limit time. Therefore, to make it practical, we assume that in the conventional strategy, the EHD is fully charged when the  energy reaches $0.99\times e_m$.} and calculate the corresponding $(E^r_i)^\prime$ before each energy harvest.
  \item[\textendash] \emph{Step 2:} Based on the sequences of $\left\{(E^h_1)^\prime,\,\ldots,\,(E^h_N)^\prime\right\}$ and $\big\{(E^r_1)^\prime,$ $\ldots,\,(E^r_{N+1})^\prime\big\}$, the length of each energy packet, $T^e_i$ for $i=1, \ldots, N$, is calculated via (\ref{eq:4_01}).
  \item[\textendash] \emph{Step 3:} Given $T^e_i$, the new offline optimal strategy and the maximal energy harvesting policy are obtained through Algorithm~\ref{alg:6_01} and (\ref{eq:4_08}), respectively.
\end{itemize}
\end{adjustwidth}
\vspace{0.1cm}
In simulations, both $(E^h_i)^\prime$ and  the length of each epoch obey a uniform distribution, the variance of which is set to one fifth of the mean value. The evaluation results are the average of $30$ independent tests.

\subsection{Performance Evaluation}
\label{subsec:perEva_onOff}

Before analyzing the simulation results, we need to highlight again that the conventional strategy violates the causality of energy harvest. According to the relationship between the harvested energy and data transmission represented in (\ref{eq:5_02}), the EHD cannot estimate the amount of energy it can harvest before scheduling its data transmission. This implies that the energy tunnel with a fixed shape that the conventional strategy assumes does not exist. Therefore, even if the conventional strategy shows a comparable performance with the new strategy in some circumstances, this strategy may not be feasible in a real system. One purpose of including the conventional strategy into comparison is to give insight into the tradeoff between the energy harvest efficiency and the energy utilization, which is discussed later.

\begin{figure}[h]
\centerline{\includegraphics[width=6.5cm]{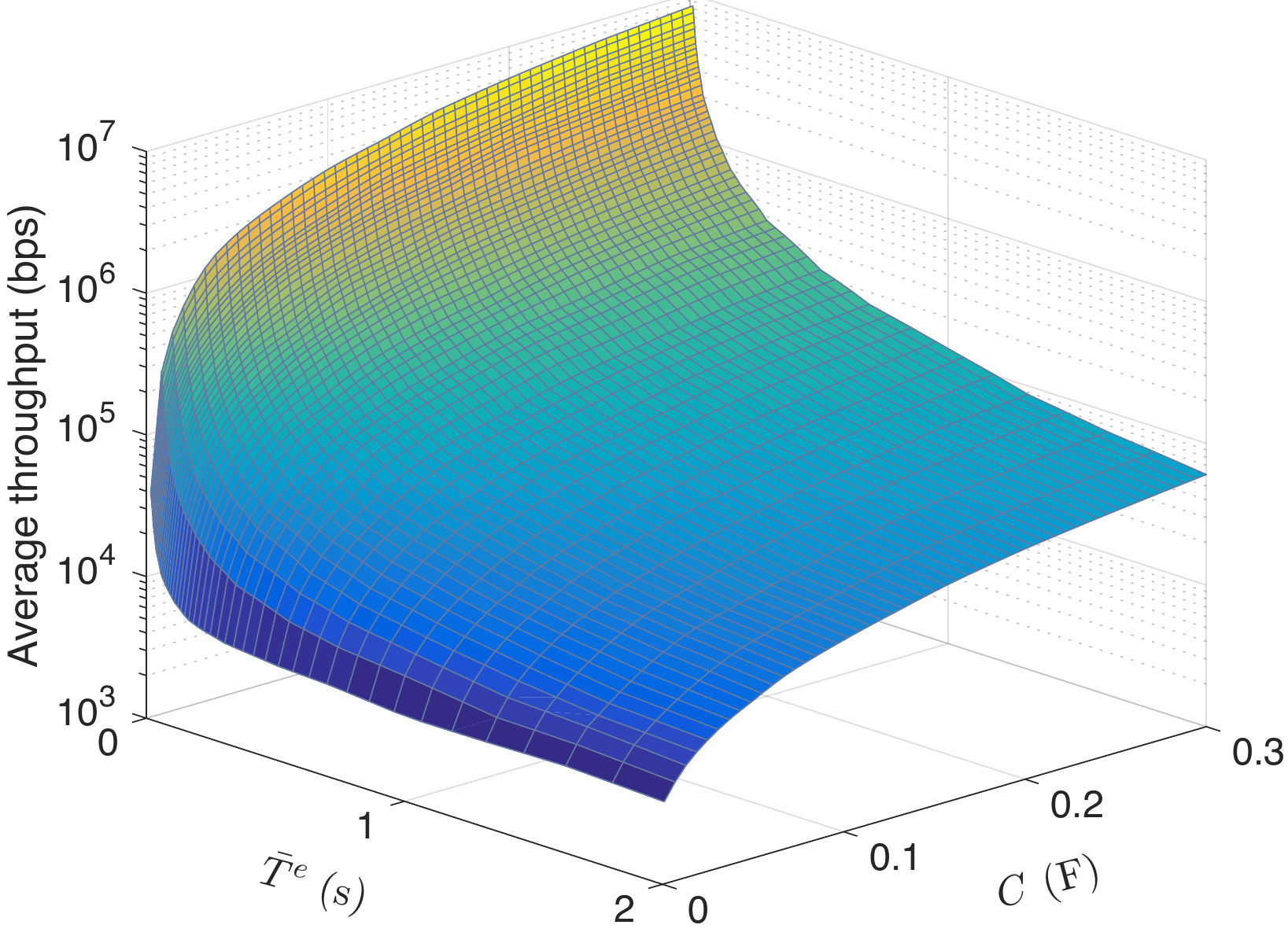}}
\caption{Throughput performance with respect to the average length of energy packets, $\bar{T^e}$, and capacitor capacity, $C$.}\label{fig:offline_para_3dlog}
\end{figure}

Fig.~\!\ref{fig:offline_para_3dlog} shows the throughput performance of the new offline optimal strategy with respect to the average length of energy packets, $\bar{T}^e$, and capacitor capacities, $C$. With the increase of $\bar{T}^e$, we reduce the frequency of energy arrivals to maintain a constant energy density, which is defined by the average amount of energy arrived at EHDs per second. It is clear that the throughput monotonously decreases with the growing length of energy packets. The main reason for this phenomenon is that the efficiency of energy harvest is reduced at the end of a long energy packet reception due to the nonlinear charge feature of EHDs discussed in Section~\ref{sec:nonlinear}. Therefore, if the energy density is a constant, EHDs prefer short energy packets to achieve a high throughput. 

To improve the efficiency on harvesting long energy packets, a supercapacitor with large capacitance can be used in the charging circuit. However, a supercapacitor with high capacity usually has a low voltage cell\footnote{Standard supercapacitors with aqueous electrolyte and organic solvents are usually specified with a rated voltage of $2.1$\,--\,$2.3$\,V and $2.5$\,--\,$2.7$\,V, respectively.} and serial connections might be required to drive the transmitter, which increases the cost and the size of the EHD. In Fig.~\!\ref{fig:offline_para_3dlog}, it can be observed that for any given $\bar{T}^e$, the improvement in throughput with respect to the increase of capacity is  logarithmic. This indicates that the rise of throughput slows down quickly with a higher $C$. For this reason, EHDs should use the supercapacitor with a proper capacitance based on the budget of a project, the length of energy packet, and the quality of service (QoS) of an application.

\begin{figure*}[htb]
\centering
\subfigure [Short energy packet]{
\label{fig:perfcmp_thr_tau:a}
\includegraphics[width=5.7cm]{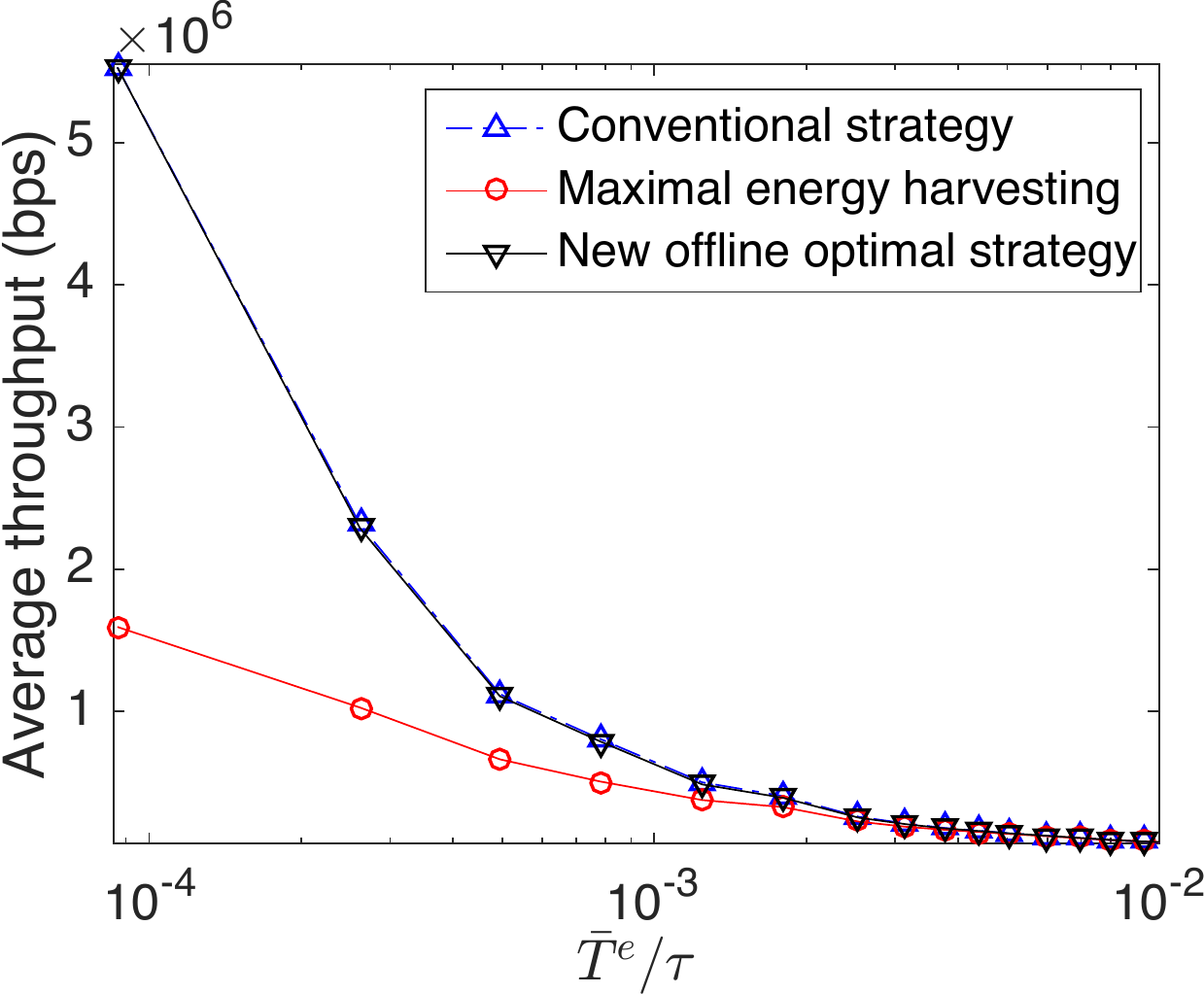}
}%
 \hspace{-0.08in}
\subfigure [Medium energy packet]{
\label{fig:perfcmp_thr_tau:b}
\includegraphics[width=5.8cm]{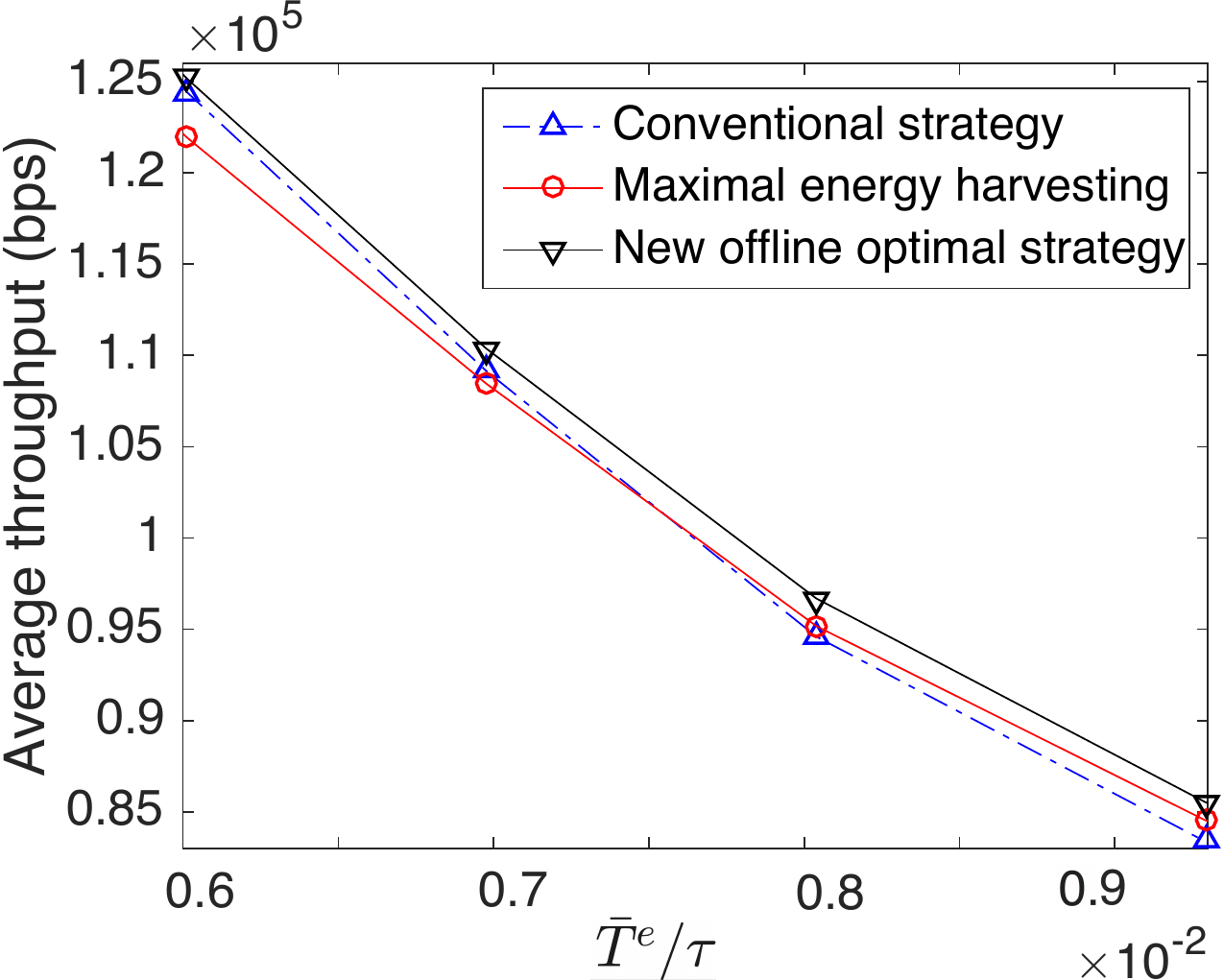}
}%
 \hspace{-0.03in}
\subfigure [Long energy packet]{
\label{fig:perfcmp_thr_tau:c}
\includegraphics[width=5.6cm]{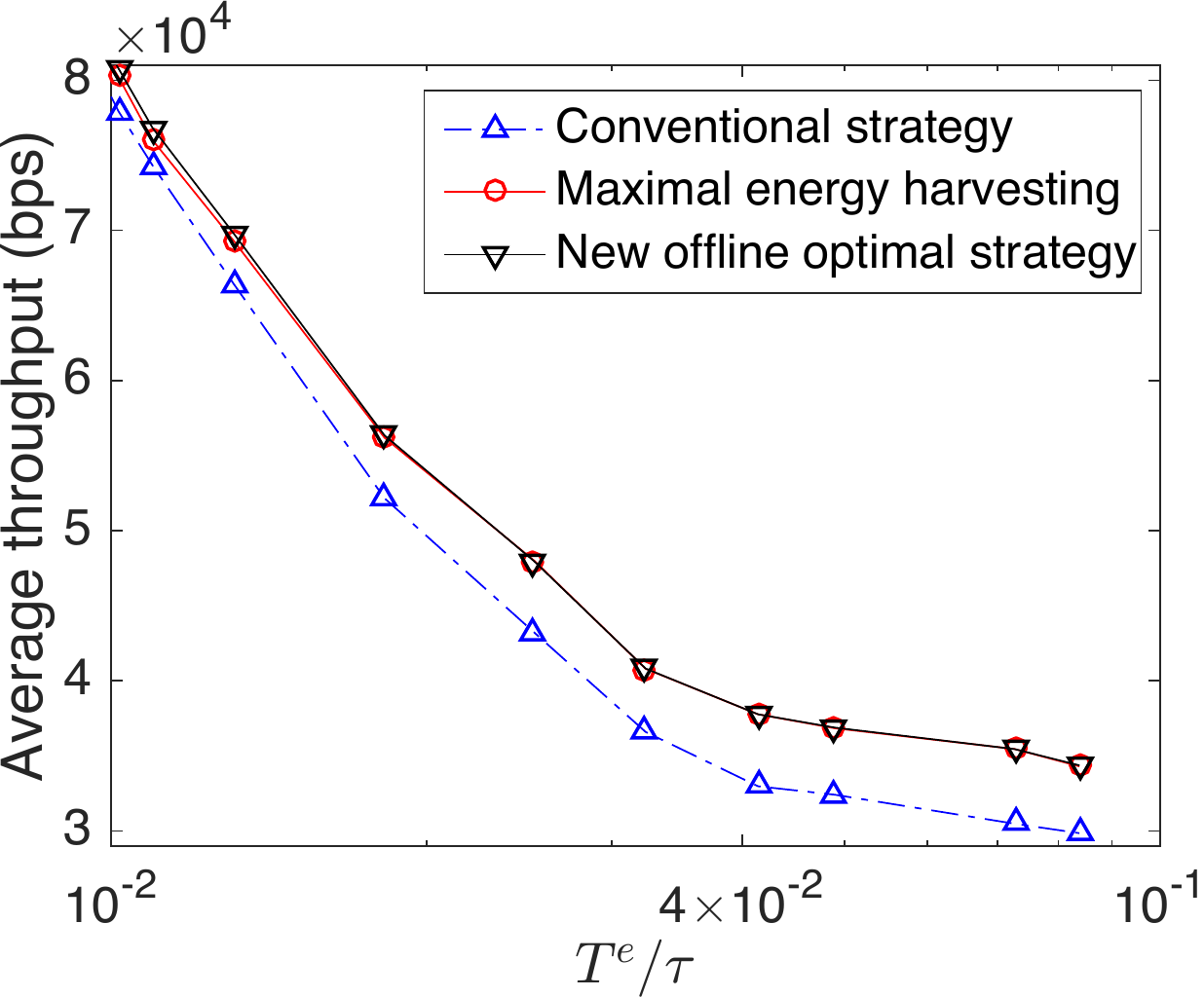}
}%
\caption{Average throughputs among different transmission strategies with respect to  $\bar{T}^e\!/\tau$.}
\label{fig:perfcmp_thr_tau}
\end{figure*}

In order to well study the throughput performance of the three strategies, we split the results into three sub-figures, where Fig.\,\ref{fig:perfcmp_thr_tau} (a), (b) and (c) displays the scenarios in average with the short length, medium length and long length of energy packets, respectively. Additionally, to eliminate the effect of the capacitor's capacitance on performance assessment, the average length of energy packets, $\bar{T}^e$, is divided by the time constant of the charging circuit, $\tau$. The throughput degradation with the growth of $\bar{T}^e\!/\tau$ is consistent to the conclusions drawn from Fig.~\ref{fig:offline_para_3dlog}. 

From Fig.\,\ref{fig:perfcmp_thr_tau} (a), it can be observed that with a small $\bar{T}^e$, the proposed strategy and the conventional one have significantly higher throughput than the maximal energy harvesting policy, which ignores the efficiency of energy utilization in the strategy design. This indicates that when the energy packets are of short length, it becomes more important to efficiently utilize the marginal energy than to improve the energy harvest efficiency. With the increase of $\bar{T}^e/\tau$ the throughput  of the maximal energy harvesting strategy gradually approaches the other two policies. As illustrated in Fig.\,\ref{fig:perfcmp_thr_tau:b}, if $\bar{T}^e/\tau$ stays in a moderate range, the new  strategy offers the highest throughput amongst the three strategies, although the advantage is not significant. As $\bar{T}^e\!/\tau$ further increases in Fig.\,\ref{fig:perfcmp_thr_tau:c}, the throughput of the new offline and the maximal energy harvesting strategies become higher than that of the conventional policy. It implies that when the RF energy is strong in the air, improving the efficiency of energy harvest brings more benefit in terms of system throughput than improving the energy utilization.

\begin{figure}[h]
\centerline{\includegraphics[width=6.5cm]{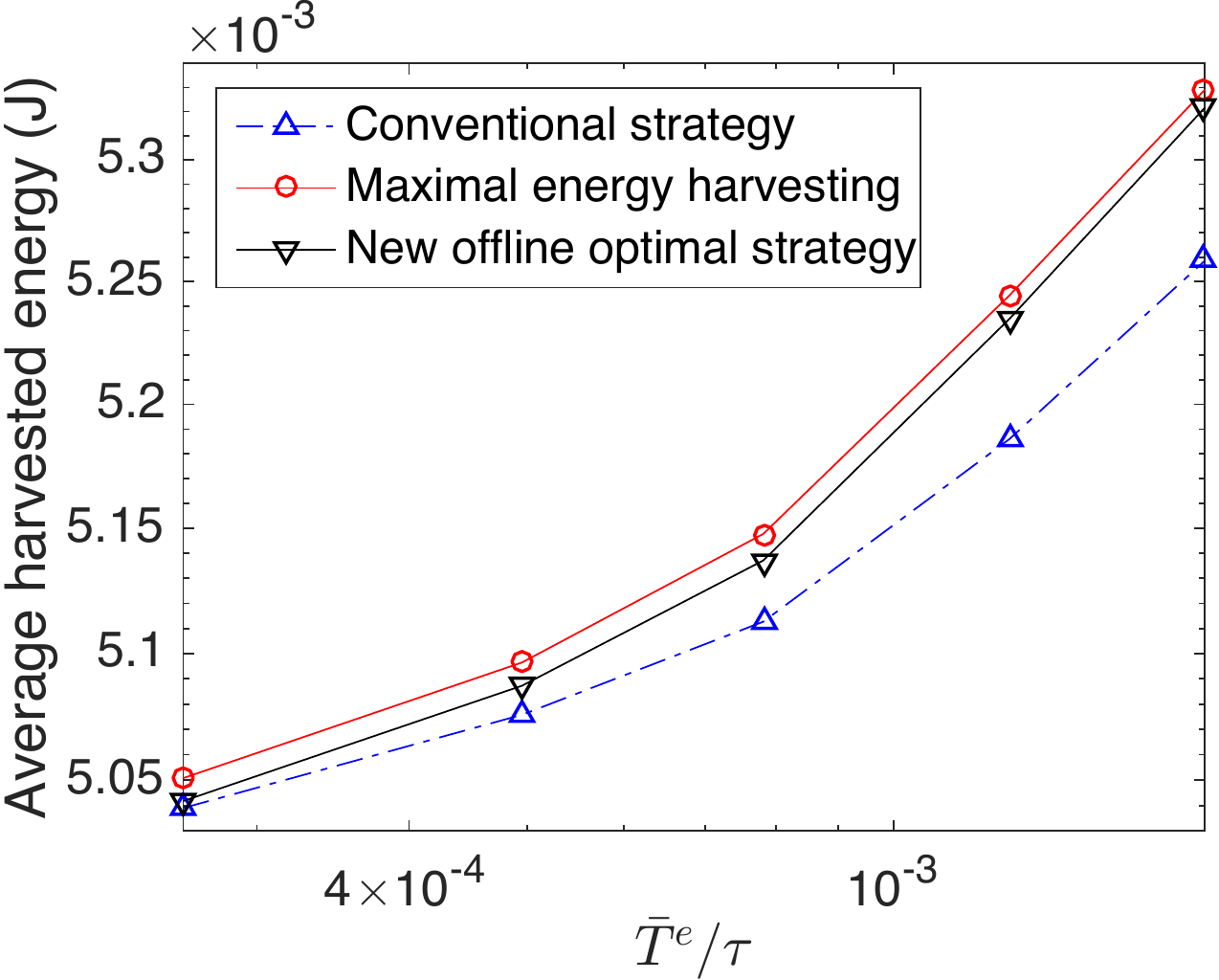}}
\caption{Harvested energy with respect to $\bar{T}^e\!/\tau$.}\label{fig:perfcmp_nrg_tau}
\end{figure}

Here, we show more details on how the efficiency of energy harvest and the energy utilization are balanced in the new transmission strategy.
In Fig.\,\ref{fig:perfcmp_nrg_tau}, we change the length of energy packets and compare the amount of energy harvested by the three transmission strategies.  As depicted in Fig.\,\ref{fig:perfcmp_nrg_tau}, the maximal energy harvesting strategy acquired the highest energy among three policies by optimizing the residual energy through (\ref{eq:4_08}). When the energy packet is short, taking $\bar{T}^e\!/\tau\!=\!2.6\!\times \!10^{-4}$ as an example, the maximal energy harvesting policy harvested $0.18\%$ more energy at the cost of $55\%$ throughput degradation in Fig.\,\ref{fig:perfcmp_thr_tau:a} compared to the new optimal strategy. When the length of energy packets increases, the difference of the amount of energy harvested by the conventional one and the other two strategies becomes significant, resulting in a low throughput as shown in Fig.\,\ref{fig:perfcmp_thr_tau:c}. This result validates that the new optimal strategy can balance the efficiencies on energy harvest and utilization dynamically based on the arrived energy. This feature makes the proposed strategy achieve better performance than the maximal energy harvesting policy, which sacrifices efficiencies on energy utilization, and the conventional strategy, which ignores the efficiency of energy harvest. 

To summarize, according to the above analysis, it could be obtained that if the energy packet is short, improving the energy utilization plays a more important role in throughput optimization compared to enhancing the energy harvest efficiency. By contrast, with long energy packets, high efficiency of energy harvest becomes a dominant factor in the design of a transmission scheduling strategy. The proposed offline optimal transmission scheduling is validated to have well balanced the energy utilization and energy harvest efficiency.

\subsection{Discussion of Online Policy}
\label{subsec:perEva_online}

Recall that Algorithm 1 reveals a property of the optimal offline strategy: the optimal transmission power of an EHD in current epoch is affected not only by the energy received before but also by future energy arrivals. However,  their impacts may be inequivalent. In particular, a nearby energy packet causes a much heavier impact on the current transmission strategy than a remote one. Therefore, if an EHD can predict the length and the arrival time of energy packets that will be received in a near future, an online strategy with near optimal performance can be expected.

\begin{figure}[htb]
\centerline{\includegraphics[width=6.5cm]{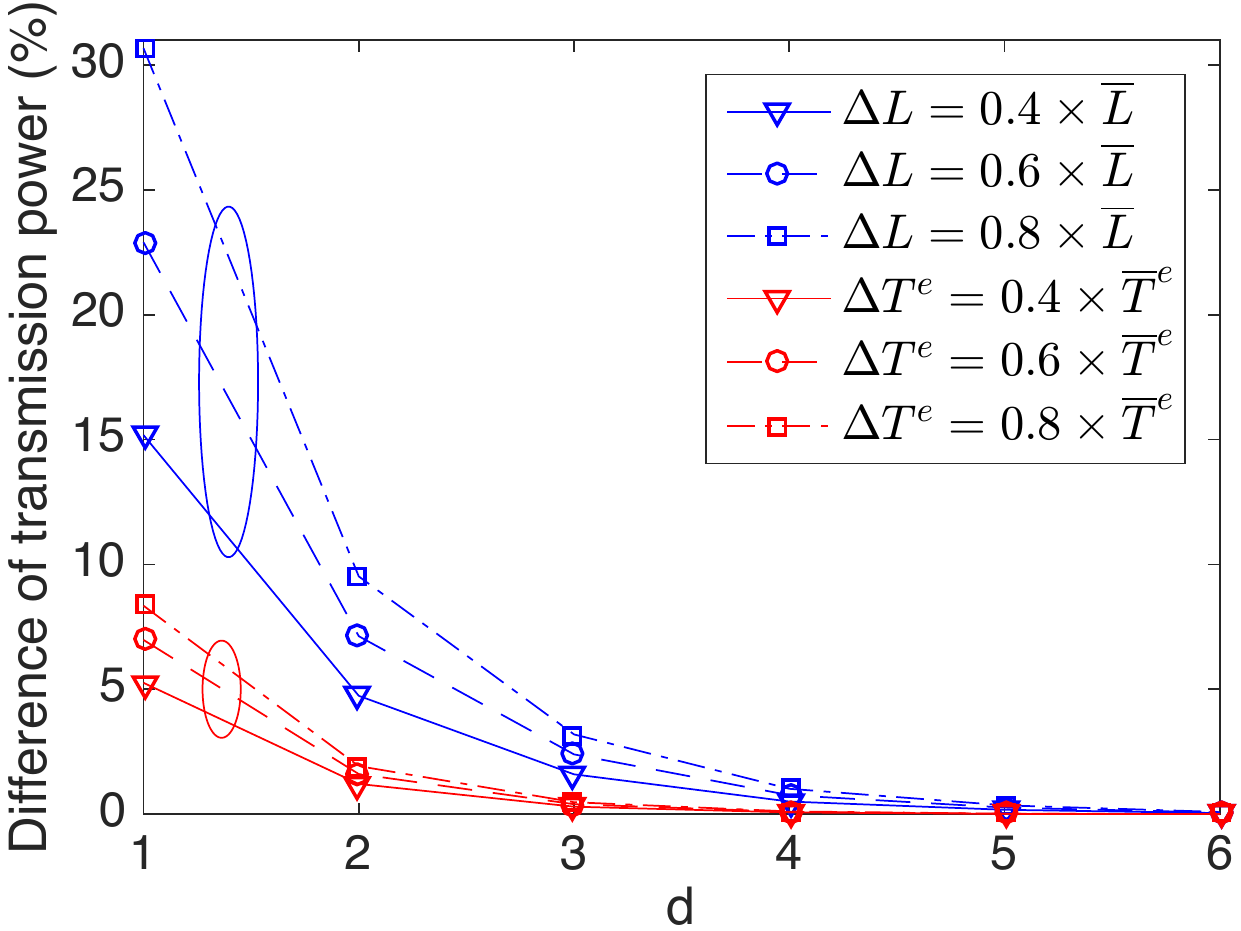}}
\caption{Impact of future energy arrivals on the transmission scheduling in the current epoch.}\label{fig:offline_time_corr}
\end{figure}

Now, we investigate the effect of future energy arrivals on the optimal transmission power in the current epoch with a medium length of energy packets. Denote the current epoch as $i$, and then the epoch that $d$ epochs way from the current one is $(i+d)$. Assume the original duration of the $(i+d)^{th}$ epoch is $\bar{L}$; the length of energy packet arrived at the $(i+d)^{th}$ epoch is $\bar{T}^e$. In Fig.\,\ref{fig:offline_time_corr}, we change $\bar{L}$ and $\bar{T}^e$ to $\bar{L}+\Delta L$ and $\bar{T}^e+\Delta \bar{T}^e$, respectively, and then present the difference of optimal transmission power before and after the variation of  $\bar{L}$ and $\bar{T}^e$.

As demonstrated in Fig.\,\ref{fig:offline_time_corr}, if the interval between energy arrivals or the length of an incoming energy packet changes, i.e., $d\!=\!1$, an EHD needs to modify its transmission power significantly so that it can better adapt to the changes of $\bar{T}^e$ and $\bar{L}$. However, if an energy packet is five epochs away, i.e., $d=5$, the impacts of varying $\bar{T}^e$ and $\bar{L}$ on the optimal transmission power in the current epoch are less than $2\%$, which is negligible in a real application. The observations from Fig.\,\ref{fig:offline_time_corr} implies that developing an online strategy with near optimal throughput is promising if an EHD can predict the time of arrival and the length of next few energy packets to some extent.

The above expectation is also validated by the simulation results of Fig.~\ref{fig:perfcmp_thr_tau}, which demonstrate the following two critical observations:

\begin{obser}
\label{obser:01}
    \emph\emph{{With medium or long energy packets (i.e., $\bar{T}^e\!/\tau\!\geq \!10^{-3}$), the optimal transmission strategy approaches the maximal energy harvesting policy.}}
\end{obser}

\begin{adjustwidth}{0cm}{0cm}
The maximal energy harvesting strategy can be considered as an online policy when the average length of energy packets is greater than $10^{-3}\tau$. The throughput can be nearly optimized by retaining a specific amount of residual energy that depends on the length of the next incoming energy packet, as presented in (\ref{eq:4_08}). Therefore, it can be implemented online with one-step prediction of energy arrivals. The length and arrival time of neighboring energy packets are usually highly correlated when the EHD is powered by a dedicated energy source, e.g.,  RF identification (RFID) system~\cite{das2014rfid, kaur2011rfid}. In this scenario, the prediction model, such as a Markov chain or an adaptive filter~\cite{michelusi2013transmission}, can be applied to estimate the information of incoming energy packet. When the energy source is ambient RF signal, the length and the time that an energy packet arrives at the EHD are random. The average length and the time interval of past energy packets can be treated as an estimation to the next energy replenishment.
\end{adjustwidth}

\begin{obser}
\label{obser:02}
    \emph\emph{{With short energy packets (i.e., $\bar{T}^e\!/\tau\!< \!10^{-3}$), the optimal transmission strategy  approaches the policy that maximizes the efficiency of energy utilization.}}
\end{obser}

\begin{adjustwidth}{0cm}{0cm}
The strategy that maximizes the efficiency of energy utilization has similar feature to existing transmission scheduling, where the battery will be drained or nearly fully charged before the change of transmission power. In this case, a state-of-the-art of the online solutions can be found in \cite{ulukus2015energy}. 
\end{adjustwidth}

%% file: Conclusion.tex
\section{Conclusions}
\label{sec:conclusion}

In this paper, a new feedback-based model has been proposed for RF energy harvesting  communications. Taking the charge characteristic of an energy harvesting circuit into account, the new model reveals the impact of data transmission on harvested energy, which introduces a new  constraint called the causality of energy harvest for the design of energy harvesting strategy. With such the constraint the feasible energy tunnel is not fixed; its bounds change with different transmission strategies dynamically. Based on the new energy harvesting model, the problem of seeking offline optimal transmission strategy has been reformulated and solved by developing a recursive algorithm. According to simulation results, the new transmission scheduling strategy is able to balance efficiencies between energy harvest and energy utilization. The design of an online policy with the new energy harvesting model is also briefly discussed. From the discussion, it can be realized that a near-optimal online strategy is available if the length and the arrival time of energy packets that will be received in the near future can be predicted in a certain level.

%% file: Appendix.tex
\appendix
\label{app:appendix}

\subsection{Proof of New Property~\ref{newpro:01}}
\label{app:appA}

\begin{proof}
We prove the New Property~\ref{newpro:01} by contradiction, assuming in an optimal transmission scheduling, the EHD's battery can be fully depleted at least once before the $(N+1)^{th}$ epoch.

Let $\texttt{ST}_\texttt{a}$ be such a strategy, which consumed all stored energy at $t_i$, $i\!\neq\!N\!+\!1$. Referring to Lemma 2 in \cite{yang2012optimal}, it can be easily proved that the optimal transmission power will not change within one epoch. Therefore, the transmission power of $\texttt{ST}_\texttt{a}$ in epochs $i$ and $i+1$ can be represented by $p_i$ and $p_{i+1}$, respectively. The overall throughput of $\texttt{ST}_\texttt{a}$ in two epochs is denoted by $Z_a$, where
\begin{equation}
\label{eq:app_01}
  Z_a = \displaystyle\frac{l_i}{2}\log_2\,(1+p_i)+\frac{l_{i+1}}{2}\!\log_2\,(1+p_{i+1}).
\end{equation}

Assume in strategy $\texttt{ST}_\texttt{b}$, the transmission power in epoch $i$ is $p_i\!-\!\Delta p_i$, where $\Delta p_i\!\in\!(0,p_i)$. Consequently, the residual energy at $t_i$ is $E^r_i\!=\!\Delta p_i\,l_i$. Denote the difference between the energy harvested in  $\texttt{ST}_\texttt{a}$ and $\texttt{ST}_\texttt{b}$ at $t_i$ by $\Delta E^h_i$, which can be calculated by substituting $E^r_i$ into (\ref{eq:4_01}), i.e.,
\begin{equation}\label{eq:app_02}
	\begin{array}{lll}
	\vspace{0.2cm}
	\Delta E^h_i\!=\!\mathcal{Q}(E^r_i)\!-\!\mathcal{Q}(0)\!=\! A^1_i(\Delta p_i\,l_i)^\frac{1}{2}\!\left[A^3_i+A^4_i(\Delta p_i\,l_i)^\frac{1}{2}\right].
	\end{array}
\end{equation}
As analyzed in Section~\ref{subsec:charge}, $A^4_i$ is negative, but $A^1_i$ and $A^3_i$ are positive values; hence we could always find a small $\Delta p_i$ that makes $\Delta E^h_i\!>\!0$. Therefore, $\texttt{ST}_\texttt{b}$ can choose higher transmission power by $\Delta p_{i+1}$, where
\begin{equation}
\label{eq:app_03}
  \Delta p_{i+1} = \displaystyle\frac{\Delta p_i\,l_i\!+\!\Delta E^h_i}{l_{i+1}}. 
\end{equation}
The overall throughput of $\texttt{ST}_\texttt{b}$ in epochs $i$ and $i\!+\!1$ will be
\begin{equation}
\label{eq:app_04}
  Z_b\!=\!\displaystyle\frac{l_i}{2}\log_2(1+p_i-\Delta p_i)+\frac{l_{i+1}}{2}\!\log_2(1+p_{i+1}\!+\!\Delta p_{i+1}).
\end{equation}

Through some simple calculations, it can be obtained  that the derivative of $Z_b\!-\!Z_a$ with respect to $\Delta p_i$ is continuous and positive infinite at $\Delta p_i \!=\!0$; meanwhile, $Z_b\!=\!Z_a$ at $\Delta p_i\!=\!0$. Accordingly, a small $\Delta p_i$ could always be found to make $Z_b\!-\!Z_a\!>\!0$, i.e.,  $Z_b\!>\!Z_a$, which indicates that $\texttt{ST}_\texttt{a}$ is not optimal. Hence, the battery cannot be fully depleted before the last epoch in an optimal scheduling strategy.
\end{proof}

\subsection{Derivation of KKT Conditions}
\label{app:appB}
To solve the stationarity equations in the KKT conditions, we need to simplify $\nabla_{\!p^*_m}\!\!\left(\sum_{j\!=\!1}^m\!\mathcal{\tilde{Q}}_j\right)$ through the following steps, where $\mathcal{\tilde{Q}}_j\!=\!E^h_j$ is the power-harvest function:
\begin{equation}\label{eq:app_06}
	\begin{array}{lll}
	\vspace{0.15cm}
	Y_m(m)\!\!\!\!\!&=&\!\!\!\! \nabla_{\!p^*_m}\!\!\left(\displaystyle\sum_{j=1}^m\mathcal{\tilde{Q}}_j\left(p^*_1,\ldots,p^*_j\right)\!\right)\\
	\vspace{0.15cm}
	&=&\!\!\!\!\nabla_{\!p^*_m}\!\!\left(\mathcal{\tilde{Q}}_m\left(p^*_1,\ldots,p^*_m\right)\right)\\
	&=&\!\!\!\!\nabla_{\!p^*_m}\!\!\left(E^r_m\right)
\!\left[\displaystyle\frac{1}{2}A^1_mA^3_m\!\left(E^r_m\right)^{-\frac{1}{2}}\!+\!A^1_mA^4_m\right].
	\end{array}
\end{equation}
Based on (\ref{eq:6_01}), we have that
\begin{equation}\label{eq:app_07}
	\begin{array}{lll}
	\vspace{0.1cm}
	\nabla_{\!p^*_m}\!\left(E^r_m\right)
\!\!\!\!\!&=&\!\!\!\!\nabla_{\!p^*_m}\!\!\left(\displaystyle\sum_{j=0}^{m-1} E^h_j- \sum_{j=1}^{m}p^*_j l_j\right)\\
	\vspace{0.0cm}
	&=&\!\!\!\!\nabla_{\!p^*_m}\!\!\left(E^h_0+\!\displaystyle\sum_{j=1}^{m-1}\!\mathcal{\tilde{Q}}_m\left(p^*_1,\ldots,p^*_j\right)\!\right)-l_m\\
	&=&\!\!\!\!-\,l_m.
	\end{array}
\end{equation}
Based on the definition of $X_i$ in (\ref{eq:app_13}), it could be obtained that
\begin{equation}\label{eq:app_08}
	\begin{array}{lll}
	\vspace{0.15cm}
	Y_m(m)\!\!\!\!\!&=&\!\!\!\!-l_m\!\left[\displaystyle\frac{1}{2}A^1_mA^3_m\left(E^r_m\right)^{-\frac{1}{2}}\!+\!A^1_mA^4_m\right]\\
	\vspace{0.1cm}
	&=&\!\!\!\! -l_mX_m=l_m\!\left[1-\left(X_m+1\right)\right].\\
	\end{array}
\end{equation}
Similarly, we have that
 \begin{equation}\label{eq:app_09}
	\begin{array}{lll}
	\vspace{0.15cm}
	Y_m(m+1)\!\!\!\!\!&=&\!\!\!\!\nabla_{\!p^*_m}\!\!\left(\displaystyle\sum_{j=1}^{m+1}E^h_j\right)\\
	\vspace{0.15cm}
	&=&\!\!\!\! \nabla_{\!p^*_m}\!\!\left(\mathcal{\tilde{Q}}_{m+1}\!\left(p^*_1,\ldots,p^*_{m+1}\right)\!\right)\!+\!Y_m(m)\\
	\vspace{0.15cm}
	&=&\!\!\!\!X_{m+1}\!\!\left[\nabla_{\!p^*_m}\!\!\left(\mathcal{\tilde{Q}}_m\left(p^*_1,\ldots,p^*_m\right)\!\right)\!-l_m\right]\!\!+\!Y_m(m)\\
	&=&\!\!\!\!l_m\left[1-\left(X_{m+1}+1\right)\left(X_m+1\right)
\right].\\
	\end{array}
\end{equation}
Eventually, $Y_m(i)$ can be represented by $X_m$ through:
\begin{equation}
\label{eq:app_10}
  Y_m(i)=l_m\!\left[1-\prod_{j=m}^i\!\left(X_j+1\right)\right].
\end{equation}

According to the properties introduced in Section\,\ref{subsec:offopt}, $E^r_i\!>\!\!0$ unless  $i\!=\!N\!+\!1$ and $E^r_i\!+\!E^h_i\!<\!e_m$. Hence in complementary slackness, we have that $\lambda_j\!=\!\mu_j\!=\!0$ for $j=1, \ldots,N$. Then
\begin{equation}\label{eq:app_11}
	\begin{array}{lll}
	\vspace{0.0cm}
	\nabla_{\!p^*_m}\!\mathcal{L}\!\!\!\!\!\!&=&\!\!\!\!\!\displaystyle\frac{l_m}{2\ln\!2\,(1\!+\!p^*_m)}\!+\!\!\lambda_{N\!+\!1}\!\!\left(\!\!\nabla_{\!p^*_m}\!\!\!\left(\sum_{j=1}^N\!\mathcal{\!\tilde{Q}}_j\!\!\left(p^*_1,\ldots,p^*_j\!\right)\!\!\right)\!\!-\!l_m\!\!\right)\\
	\vspace{0.0cm}
	&=&\!\!\!\!\displaystyle\frac{l_m}{2\ln\!2\,(1\!+\!p^*_m)}\!+\!\lambda_{N\!+\!1}\left(Y_m(N)-l_m\right)\\
	&=&\!\!\!\! l_m\!\!\left[\displaystyle\frac{1}{2\ln\!2\,(1+p^*_m)}-\lambda_{N+1}\!\prod_{j=m}^N\!\left(X_j+1\right)\right].
	\end{array}
\end{equation}
To satisfy the stationarity of KKT conditions, we have that
\begin{equation}
\label{eq:app_12}
  p^*_m=\displaystyle\frac{1}{2\lambda_{N\!+\!1}\ln\!2\displaystyle\prod_{j=m}^N\!\left(X_j+1\right)}-1,
\end{equation}
Then we can get iteration between $p^*_{m+1}$ and $p^*_{m}$ in (\ref{eq:app_13}).